\documentstyle[manuscript,aps]{revtex}
\begin{document}
\draft
 
\preprint{IMPERIAL/TP/95-96/33}
 
\title{Multiple-Scale Analysis of the Quantum Anharmonic Oscillator}

\author{Carl M. Bender\footnote{Permanent address: Department of Physics,
Washington University, St. Louis, MO 63130, USA}
and Lu\'{\i}s M. A. Bettencourt}
\address{Blackett Laboratory, Imperial College, London SW7 2BZ, United Kingdom}

\date{May 23, 1996}

\maketitle

\begin{abstract}
Conventional weak-coupling perturbation theory suffers from problems that arise
from resonant coupling of successive orders in the perturbation series.
Multiple-scale perturbation theory avoids such problems by implicitly performing
an infinite reordering and resummation of the conventional perturbation series.
Multiple-scale analysis provides a good description of the {\em classical}
anharmonic oscillator. Here, it is extended to study the Heisenberg operator
equations of motion for the quantum anharmonic oscillator. The analysis yields a
system of nonlinear operator differential equations, which is solved exactly.
The solution provides an operator mass renormalization of the theory.
\end{abstract}


Multiple-scale perturbation theory (MSPT) is a powerful and sophisticated
perturbative method for solving physical problems having a small parameter
$\epsilon$ \cite{BO,Nay}. MSPT is applicable to both linear and nonlinear
problems. Indeed, it is so general that other perturbative methods such as WKB
theory and boundary-layer theory, which are useful in more limited contexts, may
be viewed as special cases of MSPT \cite{BO}.

MSPT recognizes that dynamical systems exhibit characteristic physical behaviors
at various length or time scales. The problem with conventional perturbation
theory is that there is often a resonant coupling between successive orders.
This coupling gives rise to {\em secular terms} (terms that grow rapidly with
the length or time variable) in the perturbation series. Secular terms conflict
with physical requirements that the solution be finite. MSPT {\em reorganizes}
the conventional perturbation series to eliminate secular terms, and in doing so
it describes quantitatively the behaviors that occur at many scales.

Ordinarily, MSPT is applied to {\em classical} differential equations such as
Duffing's equation (the nonlinear equation of motion for the classical
anharmonic oscillator):
\begin{equation}
y''+y+4\epsilon y^3=0\quad (\epsilon\geq 0).
\label{e1}
\end{equation}
The positivity of $\epsilon$ ensures that $y(t)$ is bounded \cite{BO}. The
classical harmonic oscillator ($\epsilon=0$) has only one time scale, the period
of oscillation. However, when $\epsilon\neq 0$, the nonlinear term in
Eq.~(\ref{e1}) introduces many time scales. Using MSPT one can show that on a
long-time scale [$t={\rm O}(\epsilon^{-1})$] there is a frequency shift of order
$\epsilon$.

In this letter we generalize MSPT and apply it to the Heisenberg operator
equations of motion of the {\em quantum} anharmonic oscillator (the quantum
version of Duffing's equation). This generalization of MSPT techniques is
nontrivial because it gives rise to a nonlinear system of {\em operator}
differential equations \cite{OPEQNS}. We find the exact closed-form solution to
this system and thereby obtain the quantum operator analog of the classical
frequency shift -- an {\em operator mass renormalization} that expresses the
first-order shift of all energy levels.

We begin our presentation by reviewing the difficulties one encounters when one
tries to solve Duffing's equation using a conventional perturbation series,
$y(t)=\sum_{n=0}^{\infty}\epsilon^n y_n (t)$. We choose as initial conditions
\begin{equation}
y(0)=1\quad{\rm and}\quad y^{\prime}(0)=0,
\label{e2}
\end{equation}
which translate into $y_n(0)=\delta_{n,0}$ and $y^\prime_n(0)=0$, and
substitute $y(t)$ into Eq.~(\ref{e1}). To zeroth- and first-order in powers of
$\epsilon$, we have
\begin{eqnarray}
y_0''+y_0 &=& 0,
\label{e3}\\
y_1''+y_1 &=& -4y_0^3.
\label{e4}
\end {eqnarray}
The solution to Eq.~(\ref{e3}) satisfying the initial conditions is $y_0(t)=
\cos t$. Introducing this solution into Eq.~(\ref{e4}), we obtain $y_1''+y_1=
-\cos(3 t)-3\cos t$, which represents a forced harmonic oscillator whose driving
term has frequencies 3 and 1. A harmonic oscillator when driven at its natural
frequency, which in this case is 1, exhibits resonance. As a result, the
solution $y_1(t)={1\over 8}\cos(3 t)-{1\over 8}\cos t -{3\over 2}t\sin t$
contains a secular term that grows linearly with time $t$. The function $y_1(t)$
cannot be valid for long times because the exact solution to Duffing's equation
remains bounded for all $t$ \cite{BO}. Hence, the conventional perturbation
expansion is sensible only for short times $t<<\epsilon^{-1}$.

If one is clever, one can use the conventional perturbation series to determine
$y(t)$ for long times, say of order $\epsilon^{-1}$. To do so we note \cite{BO}
that the structure of the most secular (highest power in $t$) term in $y_n(t)$
has the form ${1\over 2}(3it/ 2)^n e^{it}/n! +c.c.$; we then approximate $y(t)$
by summing the most secular term in every order and the result is a cosine
function that remains bounded for all times $t$:
\begin{equation}
{1\over 2}\sum_{n=0}^{\infty}\left[\left({3i\epsilon t\over 2}\right)^n
{e^{it}\over n!}+c.c.\right]=\cos\left[\left(1+{3\over2}\epsilon\right)t\right].
\label{e5}
\end{equation}
Hence, on the long-time scale $\tau=\epsilon t$ we see a {\em frequency shift}
in the oscillator of order ${3\over 2}\epsilon$. This result is not exact
because there are less secular terms to all orders in the perturbation
expansion; such terms give rise to frequency shifts of order
$\epsilon^2,~\epsilon^3,~...~$.

The advantage of MSPT is that it reproduces Eq.~(\ref{e5}) directly and bypasses
the elaborate procedure of summing the conventional perturbation series to all
orders by excluding {\em ab initio} secular terms from the perturbation
expansion. MSPT assumes the existence of many time scales ($t$, $\tau=\epsilon
t$, $\sigma= \epsilon^2 t$, $...$), which can be temporarily treated as {\em
independent} variables. Here, we illustrate by performing just a first-order
calculation. We use only the two variables $t$ and $\tau=\epsilon t$ and seek a
perturbative solution to Eq.~(\ref{e1}) of the form
\begin{equation}
y(t)=Y_0(t,\tau)+\epsilon Y_1(t,\tau)+{\rm O}(\epsilon^2).
\label{e6}
\end{equation}

The chain rule and the identity ${d\tau\over dt}=\epsilon$ convert
Eq.~(\ref{e1}) to a sequence of {\em partial} differential equations for the
dependent variables $Y_0$, $Y_1$, $...~$. The first two are
\begin{eqnarray}
{\partial^2\over \partial t^2}Y_0+Y_0 &=& 0,
\label{e7} \\
{\partial^2\over \partial t^2}Y_1+Y_1 &=& -4Y_0^3-2 
{\partial^2\over\partial t\partial\tau}Y_0.
\label{e8}
\end {eqnarray}

The general solution to Eq.~(\ref{e7}) is $Y_0 (t,\tau)=A(\tau)\cos t +B(\tau)
\sin t$. We substitute $Y_0(t,\tau)$ into the right side of Eq.~(\ref{e8}) and
use triple-angle formulas such as $\cos^3 t={1\over 4}\cos(3 t)+{3\over 4}
\cos t$ to simplify the result. To determine the functions $A(\tau)$ and
$B(\tau)$ we demand that there be no resonant coupling between zeroth and first
order in perturbation theory so that no secular terms appear in $Y_1$. That is,
we require that the coefficients of $\sin t$ and $\cos t$ vanish:
\begin{equation}
2{dB\over d\tau}=-3A^3-3AB^2\quad{\rm and}\quad 2{dA\over d\tau}=3B^3+3A^2B.
\label{e9}
\end{equation}

To solve this system we multiply the first equation by $B(\tau)$, the second
by $A(\tau)$, and add the resulting equations. Letting $C(\tau)={1\over 2}
\left[ A(\tau)\right]^2+{1\over 2}\left[ B(\tau)\right]^2$, we obtain
\begin{equation}
{d\over d\tau}C(\tau)=0.
\label{e10}
\end{equation}
Thus, $C(\tau)$ is the constant $C(0)$ and the differential equation system
(\ref{e9}) becomes linear:
\begin{eqnarray}
{d\over d\tau}B=-3C(0)A \quad {\rm and} \quad {d\over d\tau}A=3C(0)B.
\label{e11}
\end{eqnarray}
The solution to this system that satisfies the initial conditions is $C(0)={1
\over 2}$ and $Y_0(t,\tau)=\cos\left[(1+{3\over 2}\epsilon) t \right]$, which
reproduces the approximate solution in Eq.(\ref{e5}). While conventional
perturbation theory is valid for $t<<\epsilon^{-1}$, the MSPT approximation is
valid for $t<<\epsilon^{-2}$.

We now consider the {\em quantum} anharmonic oscillator, whose Hamiltonian is
$H(p,q)={1\over 2}p^2+{1\over 2}q^2+\epsilon q^4$. Here, $\epsilon\geq 0$ so
that the spectrum of $H(p,q)$ is bounded below and $p$ and $q$ are operators
satisfying the canonical equal-time commutation relation $[q(t),p(t)]=i\hbar$.

The Heisenberg operator equations of motion, ${d\over dt}q ={1\over i\hbar}
[q,H(p,q)]=p$ and ${d\over dt}p={1\over i\hbar}[p,H(p,q)]=-q-4\epsilon q^3$,
combine to give the quantum Duffing's equation (\ref{e1}):
\begin{eqnarray}
{d^2\over dt^2}q+q+4\epsilon q^3=0.
\label{e12} 
\end{eqnarray}
Since $p(t)$ and $q(t)$ are operators, we cannot impose numerical initial
conditions like those in Eq.~(\ref{e2}); rather, we enforce a general operator
initial condition at $t=0$:
\begin{equation}
q(0) = q_0 \quad {\rm and}\quad p(0) = p_0,
\label{e13}
\end{equation}  
where $p_0$ and $q_0$ are time-independent operators obeying the Heisenberg
algebra $[q_0,p_0]=i\hbar$. 

We now apply MSPT to Eq.~(\ref{e12}). Assuming that $q(t)$ exhibits
characteristic behavior on the short-time scale $t$ and on the long-time scale
$\tau=\epsilon t$, we write
\begin{equation}
q(t)=Q(t,\tau)=Q_0(t,\tau)+\epsilon Q_1(t,\tau)+{\rm O}(\epsilon^2).
\label{e14}
\end{equation}
This equation is analogous to Eq.~(\ref{e6}) but here $Q_0$ and $Q_1$ are
{\em operator-valued} functions.

We substitute $q(t)$ in Eq.~(\ref{e14}) into Eq.~(\ref{e12}), collect the
coefficients of $\epsilon^0$ and $\epsilon^1$, and obtain operator differential
equations analogous to Eqs.~(\ref{e7}) and (\ref{e8}):
\begin{eqnarray}
{\partial^2\over\partial t^2}Q_0+Q_0 &=& 0,\label{e15}\\
{\partial^2\over\partial t^2}Q_1+Q_1 &=& -4Q_0^3-2{\partial^2\over
\partial t\partial\tau}Q_0.
\label{e16}
\end{eqnarray}
Because Eq.~(\ref{e15}) is linear, it is easy to find its general solution,
\begin{eqnarray}
Q_0(t,\tau)={\cal A}(\tau)\cos t +{\cal B}(\tau)\sin t,
\label{e17}
\end{eqnarray} 
and from $p={dq\over dt}$ we obtain the momentum operator
$p(t)={\cal B}(\tau)\cos t -{\cal A}(\tau)\sin t +{\rm O}(\epsilon)$.

It is now necessary to find the coefficient functions ${\cal A}(\tau)$ and
${\cal B}(\tau)$, which are operators. The canonical commutation relation in
$[q(t),p(t)]=i\hbar$ implies that these operators satisfy $[{\cal A}(\tau),
{\cal B}(\tau)]=i\hbar$. Also, the initial conditions in Eq.~(\ref{e13}) give
\begin{equation}
{\cal A}(0) = q_0 \quad {\rm and}\quad {\cal B}(0) = p_0.
\label{e18}
\end{equation}

We determine ${\cal A}(\tau)$ and ${\cal B}(\tau)$ by evaluating the right side
of Eq.~(\ref{e16}) and expanding the cubic term, taking care to preserve the
order of operator multiplication. As in the classical case, we simplify the
result using triple-angle formulas. To prevent secularity in $Q_1(t,\tau)$ we
set the coefficients of $\cos t$ and $\sin t$ to zero and obtain
\begin{eqnarray}
2{d{\cal B}\over d\tau} &=& -3{\cal A}^3-{\cal B}{\cal A}{\cal B}
-{\cal B}{\cal B}{\cal A}-{\cal A}{\cal B}{\cal B},\nonumber\\
2{d{\cal A}\over d\tau} &=& 3{\cal B}^3+{\cal A}{\cal B}{\cal A}+{\cal A}
{\cal A}{\cal B}+{\cal B}{\cal A}{\cal A}.
\label{e19}
\end{eqnarray}
This system of operator-valued differential equations is the quantum analog of
Eq.~(\ref{e9}).

To solve the system (\ref{e19}) we pre- and post-multiply the first equation
by ${\cal B}(\tau)$ and the second equation by ${\cal A}(\tau)$. Adding the
resulting four equations, we get
\begin{equation}
{d\over d\tau}{\cal H}=0,
\label{e20}
\end{equation}
where ${\cal H}\equiv {1\over 2}{\cal A}^2+{1\over 2}{\cal B}^2$. Equation
(\ref{e20}), the quantum analog of Eq.~(\ref{e10}), shows that ${\cal H}$ is
independent of the long-time variable $\tau$. Thus, Eq.~(\ref{e18}) allows us to
express ${\cal H}$ in terms of the fundamental operators $p_0$ and $q_0$:
${\cal H}={1\over 2}p_0^2+{1\over 2}q_0^2$. We then use the commutator
$[{\cal A}(\tau),{\cal B}(\tau)]=i\hbar$ to rewrite Eq.~(\ref{e19}) in
manifestly Hermitian form:
\begin{eqnarray}
{d\over d\tau}{\cal B}=-{3\over 2}\left({\cal H}{\cal A}+{\cal A}{\cal H}
\right), \quad {d\over d\tau}{\cal A}={3\over 2}\left({\cal H}{\cal B}
+{\cal B}{\cal H}\right).
\label{e21}
\end{eqnarray}

Suppose for a moment that we could replace the operator ${\cal H}$ by the
numerical constant $C(0)$ in Eq.~(\ref{e21}). Then we would obtain the
$c$-number coupled differential equations in Eq.~(\ref{e11}). That system is
linear so we could ignore operator ordering and easily find the solution
satisfying the initial conditions (\ref{e18}): ${\cal A}(\tau)=q_0\cos[3C(0)
\tau]+p_0\sin[3C(0)\tau]$ and ${\cal B}(\tau)=p_0\cos[3C(0)\tau]-q_0\sin[3C(0)
\tau]$. This solution suggests the structure of the exact solution to the
{\em operator} differential equation system (\ref{e21}). The formal solution
is a natural generalization using Weyl-ordered products of operators:
\begin{eqnarray}
{\cal A}(\tau) &=& {\cal W}[q_0\cos(3{\cal H}\tau)+p_0\sin(3{\cal H}\tau)],
\nonumber\\
{\cal B}(\tau) &=& {\cal W}[p_0\cos(3{\cal H}\tau)-q_0\sin(3{\cal H}\tau)].
\label{e22}
\end{eqnarray}

The operator ordering ${\cal W}[q_0 f({\cal H}\tau)]$ is defined as follows: (1)
Expand $f({\cal H}\tau)$ as a Taylor series in powers of the operator ${\cal H}
\tau$; (2) Weyl-order the Taylor series term-by-term: ${\cal W}(q_0{\cal H}^n)
\equiv{1\over 2^n}\sum_{j=0}^{n}({^n_j}){\cal H}^j q_0 {\cal H}^{n-j}$. Using
this definition it is easy to verify that Eq.~(\ref{e22}) is indeed the
{\em exact operator solution} to Eq.~(\ref{e21}) satisfying the initial
conditions (\ref{e18}). We simplify the formal solution in Eq.~(\ref{e22})
and re-express it in closed form by observing that if we reorder ${\cal W}(q_0{
\cal H}^n)$ by commuting $q_0$ symmetrically to the left and to the right to
maintain the Hermitian form we generate a set of polynomials \cite{POLY} of
degree $n$:
\begin{eqnarray}
{\cal W}(q_0{\cal H}^n)={\hbar^n\over 2}\left[q_0 E_n\left({{\cal H}\over\hbar}
+{1\over 2}\right)+E_n\left({{\cal H}\over\hbar}+{1\over 2}\right) q_0 \right],
\nonumber
\end{eqnarray}
We identify $E_n$ as the $n$th Euler polynomial \cite{AS} in which the argument
is shifted by $1\over 2$:
\begin{eqnarray}
1,\quad {{\cal H}\over\hbar},\quad {{\cal H}^2\over\hbar^2}-{1\over 4},\quad
{{\cal H}^3\over\hbar^3}-{3\over 4} {{\cal H}\over\hbar},\quad {{\cal H}^4\over
\hbar^4}-{3\over 2}{{\cal H}^2\over\hbar^2}+{5\over 16}, 
\nonumber
\end{eqnarray} 
and so on. The generating function for these nonorthogonal polynomials is
\begin{eqnarray}
{2e^{\left({{\cal H}\over\hbar}+{1\over 2}\right)\tau}\over e^\tau+1}=
\sum_{n=0}^{\infty}{\tau^n \over n!}E_n\left({{\cal H}\over\hbar}+{1\over 2}
\right)\quad (|\tau|<\pi).
\nonumber
\end{eqnarray}
This generating function allows us to express the following Weyl-ordered product
compactly:
\begin{equation}
{\cal W}\left(q_0 e^{{\cal H}\tau}\right)=
{q_0 e^{{\cal H}\tau}+e^{{\cal H}\tau}q_0\over 2\cosh (\tau\hbar/2)}.
\label{e23}
\end{equation}

Using Eq.~(\ref{e23}) we rewrite compactly the cosines and sines in
Eq.~(\ref{e22}), substitute this result into Eq.~(\ref{e17}), replace $\tau$ by
$\epsilon t$, and obtain
\begin{eqnarray}
Q_0(t,\tau) &=& {q_0\cos(t+3{\cal H}\epsilon t)+\cos(t+3{\cal H}\epsilon t)q_0
\over 2\cos (3\epsilon t\hbar/2)}\nonumber\\
&&\quad +{p_0\sin(t+3{\cal H}\epsilon t)+\sin(t+3{\cal H}
\epsilon t)p_0\over 2\cos (3\epsilon t\hbar/2)}.
\label{e24}
\end{eqnarray}
Equation (\ref{e24}) is the quantum operator analog of Eq.~(\ref{e5}) and is
the objective of our multiscale analysis. [We can recover the classical MSPT
approximation in Eq.~(\ref{e5}) by taking the limit $\hbar\to 0$ and imposing
the classical initial conditions $p_0=0$ and $q_0=1$, which give ${\cal H}={1
\over 2}$.] Recall that in the classical case we identify the coefficient of the
time $t$ as a first-order approximation to the frequency shift. Since the
coefficient of $t$ in Eq.~(\ref{e24}) is an operator, we have derived an
{\em operator form of mass renormalization}.

To elucidate this operator mass renormalization we study matrix elements of
Eq.~(\ref{e24}). The time dependence of a matrix element reveals the
energy-level differences of the quantum system. It is easy to construct a set of
states because the operators $q_0$ and $p_0$ satisfy the Heisenberg algebra
$[q_0, p_0]=i\hbar$. Hence, appropriate linear combinations of $q_0$ and $p_0$
may be used as raising and lowering operators to generate a Fock space
consisting of the states $| n\rangle$. By construction, these states are
eigenstates of the operator ${\cal H}$: ${\cal H}|n\rangle=\left( n+{1\over 2}
\right)\hbar |n\rangle$. Evaluating Eq.~(\ref{e24}) between the states
$\langle n-1|$ and $| n\rangle$ and allowing the operator ${\cal H}$ to act to
the left and the right, we obtain
\begin{eqnarray}
\langle n-1| Q_0 | n\rangle &=& \langle n-1| q_0 | n\rangle
\cos [t(1+3n\hbar\epsilon )]\nonumber\\
&&\quad +\langle n-1| p_0 | n\rangle \sin [t(1+3n\hbar\epsilon )],
\label{e25}
\end{eqnarray}
which predicts that the energy-level differences of the quantum oscillator are
$1+3n\hbar\epsilon$. We may verify this result by recalling that the first-order
correction to the energy eigenvalues \cite{BW} is $E_n=n+{1\over 2}+{3\over 4}
\epsilon\hbar (2n^2+2n+1)+{\rm O}(\epsilon^2)$. Thus, $E_n-E_{n-1}=1+3n\hbar
\epsilon+{\rm O} (\epsilon^2)$.

We conclude by noting that in addition to the operators $q(t)$ and $p(t)$, the
wave function $\psi(x)$ for the quantum anharmonic oscillator also exhibits
multiscale behavior. Specifically, to all orders in conventional weak-coupling
Rayleigh-Schr\"odinger perturbation theory, $\psi(x)$ behaves like the Gaussian
$e^{-x^2/4}$ for large $x$; however, a geometrical-optics approximation to
$\psi(x)$ from WKB theory predicts that for large $x$, $\psi(x)$ decays like the
exponential of a cubic $e^{-\sqrt{\epsilon}\vert x\vert^3/6}$. We can resolve
this discrepancy by using a MSPT approach; we reorder the perturbation series by
resumming secular terms \cite{BB}.

The wave function $\psi(x)$ obeys the Schr\"odinger equation
\begin{eqnarray}
\left( -{d^2 \over dx^2} + {1\over 4}x^2 + {1\over 4}\epsilon x^4
-E(\epsilon) \right)\psi(x) = 0
\label{e26}
\end{eqnarray}
and satisfies $\psi(\pm\infty)=0$. The conventional perturbative approach to
Eq.~(\ref{e26}) \cite{BW} represents both the eigenfunction and eigenvalue
as asymptotic series in $\epsilon$: $\psi(x)\sim\sum_{n=0}^\infty\epsilon^n y_n
(x)$ and $E(\epsilon)\sim\sum_{n=0}^\infty\epsilon^n E_n$. In Ref.~\cite{BW} it
is shown that for the ground state, $y_n(x)$ is a Gaussian multiplied by a
polynomial of degree $2n$ in the variable $x^2$, $y_n=e^{-x^2/4}P_n(x)$, where
\begin{eqnarray}
P_0(x)=1\quad {\rm and}\quad
P_n(x)=\sum_{j=1}^{2 n} C_{n,k} (-{1\over 2}x^2)^k \quad (n>0).
\nonumber
\end{eqnarray}
The recursion relation for the polynomials $P_n(x)$ is typical of all
perturbative calculations; the homogeneous part of this recursion relation is
independent of $n$ while the inhomogeneous part contains all previous
polynomials. This recursive structure is responsible for successive orders of
perturbation theory being resonantly coupled and causes the degree of the
polynomials to grow with $n$.

To study the behavior of the wave function $\psi(x)$ for large $x$ we
approximate $\psi(x)$ by resumming the perturbation series and keeping just the
{\em highest power} of $x$ in every order. [This is an exact analog of finding
the coefficient of the highest power of $t$ (most secular term) in $n$th order
in perturbation theory for the classical anharmonic oscillator.] This
resummation gives a new representation of $\psi(x)$: $e^{-x^2/4}e^{-\epsilon x^4
/16}$ multiplied by a {\em new set of polynomials}. For the classical anharmonic
oscillator summing leading secular terms also gives an exponential approximation
[see Eq.~(\ref{e5})]. However, the classical and quantum anharmonic oscillators
are quite different; although we have summed the most secular terms to all
orders in perturbation theory, the result is {\em not} the actual behavior of
the wave function $\psi(x)$ for large $x$; the correct behavior is an
exponential of a cubic and not a quartic!

If we iterate this resummation process we find that each reorganization of the
perturbation series gives an additional term in the exponential
\begin{eqnarray}
{\rm exp}\left(-{1\over 4}x^2-{1\over 16}\epsilon x^4+{1\over 96}\epsilon^2x^6
-{1\over 256}\epsilon^3 x^8+...\right)
\label{e27}
\end{eqnarray}
and a new set of polynomials. It seems impossible for this approach to give
cubic exponential behavior because at each stage in the reorganization of the
perturbation series the variable $x$ appears only in {\em even} powers. However,
we recognize that the exponent in Eq.~(\ref{e27}) is the beginning of a binomial
series whose sum is ${1\over 6\epsilon}\left[ 1-(1+\epsilon x^2)^{3/2}\right]$.
If we now let $x$ be large $(\epsilon x^2>>1)$, we recover the correct
asymptotic behavior of $\psi(x)$ \cite{GM}.

The approach used above for the anharmonic oscillator wave function has been 
used in perturbative quantum field theory to sum leading-logarithm divergences
\cite{CW} and leading infrared divergences \cite{DJP}. It is our hope that in
the future the direct nonperturbative multivariate approach of MSPT will provide
a framework to simplify such schemes.
 
CMB thanks the Department of Theoretical Physics at Imperial College, London,
for its hospitality and the Fulbright Foundation, the PPARC, and the U.S.
Department of Energy for financial support. LMAB thanks JNICT--{\it Programa
Praxis XXI} for financial support under contract BD 2243/92.

\end{document}